\documentclass[12pt]{article}

\newcommand{\be}{\begin{equation}}
\newcommand{\bea}{\begin{eqnarray}}
\newcommand{\ee}{\end{equation}}
\newcommand{\eea}{\end{eqnarray}}

\def\theequation{\arabic{section}.\arabic{equation}}
\begin{document}
\topmargin -1cm \oddsidemargin=0.25cm\evensidemargin=0.25cm
\setcounter{page}0
\renewcommand{\thefootnote}{\fnsymbol{footnote}}
\begin{titlepage}
\begin{flushright}
LTH--746 \\
\end{flushright}
\vskip .7in
\begin{center}
{\Large \bf On the
Low Energy Spectra of the  Nonsupersymmetric Heterotic String
Theories} \vskip .7in
{\large Alon E. Faraggi}$^a$
and
{\large Mirian Tsulaia$^b$}
 \vskip .2in {$^a$ \it
Department of Mathematical Sciences,
University of Liverpool,
Liverpool L69 7ZL, UK} \\
\vskip .2in {$^b$ \it
Department of Physics and Institute of Plasma Physics,
University of Crete, 71003 Heraklion, Crete, Greece} \\

\begin{abstract}

The ten dimensional string theories as well as eleven dimensional
supergravity are conjectured to arise as limits of a more basic
theory, traditionally dubbed M--theory. This notion is confined to
the ten dimensional supersymmetric theories. String theory, however,
also contains ten dimensional non--supersymmetric theories that have
not been incorporated into this picture. In this note we explore the
possibility of generating the low energy spectra of various
nonsupersymmetric heterotic string vacua from the Horava--Witten
model. We argue that this can be achieved by imposing on the
Horava--Witten model an invariance with respect to some extra
operators which identify the orbifold fixed planes in a nontrivial
way, and demonstrate it for the $E_8$ and $SO(16)\times SO(16)$
heterotic string vacua in ten dimensions.

\end{abstract}

\end{center}

\vfill

\end{titlepage}


\setcounter{equation}0\section{Introduction}

Superstring theory provides a consistent framework to explore the
unification of gravity with the gauge interactions. The main feature
of string theory is that it maintains the interpretation of the
fundamental constituents of matter as elementary particles with
internal quantum attributes. The string view of elementary particles
is  therefore a natural extension of the point particle
interpretation, and may be regarded as not distinct from it. The
internal consistency of string theory imposes restrictions on the
possible internal attributes. String theory facilitates the
development of a phenomenological approach to quantum gravity, which
is constrained by the string self--consistency requirements.

The internal consistency of string theory introduces the possibility
of interpreting a number of degrees of freedom, required for consistency,
as bosonic extra space--time dimensions. Thus, the bosonic string
is formulated in twenty--six dimensions, while the fermionic strings
are formulated in ten space time dimensions. An important step in
the development of string theory
was obtained by the understanding that the five different
supersymmetric ten dimensional theories may be related by perturbative
and nonperturbative duality transformations
\cite{Hull:1994ys,Witten:1995ex}.
It was further suggested
that the ten dimensional string theories can be related to eleven
dimensional supergravity by compactification on $S_1$ and $S_1/Z_2$
\cite{Duff:1987bx,Horava:1995qa,Horava:1996ma}.
In this regard the duality picture provides a compelling understanding
of the ten dimensional string theories and indicates the existence
of a more fundamental underlying structure.

The problem of supersymmetry breaking, and its realization in
nature, is one of the vital issues in string theory. In addition to
the supersymmetric ten dimensional fermionic string theories there
exist ten dimensional fermionic string theories that are
non--supersymmetric \cite{Kawai:1986vd,Dixon:1986iz}. These vacua
have not been incorporated into the ten dimensional duality picture,
although some preliminary studies were carried out in
\cite{Blum:1997gw, Bergman:1997rf}. It is likely that progress on
understanding how to incorporate these non--supersymmetric vacua in
the ten dimensional duality picture
(see \cite{Aharony:2007du} for a recent discussion in the
framework of the M(atrix) theory), may be instrumental to progress
on the problem of supersymmetry breaking in string theory.

In this paper we make an attempt in this direction. Our basic
starting point is the eleven dimensional supergravity field theory
compactified on $S_1/Z_2$, \`a la Horava--Witten. We then study the
possibility of reproducing the field theory content of the ten
dimensional non--supersymmetric heterotic strings by imposing some
extra symmetry operation on the the Horava--Witten model. The
basic ingredient in our analysis is the gluing of the end points
of the Horava--Witten theory at $x^{10}=0$ and $x^{10}= \pi$.
However, the picture is
further complicated due to the existence of the branes and the
fields that reside on them.
The essential point in our construction is an identification of the
fields present on the boundary branes, located at the
orbifold fixed points $x^{10}=0$ and $x^{10}= \pi$.
This operation is augmented
by an extra operation which interchanges the gauge degrees of freedom,
in complete analogy with the one discussed in ref. \cite{Dixon:1986iz}.
Identification of the degrees of freedom on the two
different branes, located at the different orbifold fixed points,
leads in turn to the identification of the branes themselves in
order to maintain the localty of the low energy theory. The
massless spectrum of the resulting
models is obtained from the
massless spectrum of the Horava--Witten model, by taking the
$x^{10}$ coordinate to be compact. In other words
the operation we describe puts the orbifold fixed planes on top of
each other with additional identification of the gauge degrees of
freedom.  This operation projects out certain fields from
the initial spectrum (for example it projects out fermions form
supergravity sector in the first model) and in order to cancel
the anomalies one has to add
fermions with certain chirality on the resulting brane, in
analogy with the Horava--Witten construction.
In this manner we reproduce the massless spectrum of various
nonsupersymmetric ten dimensional
heterotic string vacua from the
Horava--Witten model, by the operation which
essentially changes the geometry of the later.

In the first example we investigate how to
reproduce the spectrum of the $E_8$ Kac--Moody level two model. We
propose that this can be achieved by imposing that the two fixed
branes of the Horava--Witten model are identified by the orbifold
condition, and imposing that the $E_8\otimes E_8$ gauge group is
broken to the diagonal $E_8$ level two subgroup. Anomaly
cancellation is satisfied by adding a set of fermion fields from the
twisted sector, as in the ten dimensional  heterotic string vacuum.
In the second example we discuss a model where the entire part of
$N=1$ $D=10$ supergravity multiplet is kept intact. As for the
previous example, the model is anomalous unless we add extra
massless fermions to the spectrum. In the final example we impose
some extra operators onto the spectrum of Horava--Witten model to
obtain the massless spectrum of
 the nonsupersymmetric $SO(16) \otimes SO(16)$
heterotic string model
in a way similar to the previous two models.

\setcounter{equation}0\section{Basic Definitions}

\subsection{11 Dimensional Supergravity on $S_1/Z_2$}

First let us collect the results of
\cite{Horava:1995qa}--\cite{Horava:1996ma}. This system corresponds
to the effective theory of the strongly coupled $E_8 \otimes E_8$
heterotic superstring. It is described by the eleven dimensional
supergravity when  one dimension, $x^{10}$, is compactified on
 $S^1/Z_2$ orbifold.
 In this construction there are two  ten dimensional ``mirror``
 branes located at the orbifold fixed points $x^{10}=0$ and $x^{10}= \pi$.
The field content in the eleven --dimensional bulk is the one of the
usual eleven dimensional supergravity i.e., it consists of the
graviton $G_{MN}$, gravitino $\Psi_M$ and antisymmetric tensor field
of the third rank $C_{MNP}$, $M,N,P = 0, \dots, 10$. In order to
obtain the field content on the orbifold fixed planes one assigns
various eigenvalues to the fields under  consideration with respect
to the $Z_2$ parity operator ${\cal R}$. We have the following
action of the parity operator on bosons
\begin{equation} \label{BP}
{\cal R}G_{\mu \nu} = G_{\mu \nu}, \quad {\cal R}C_{\mu \nu  \rho} = - C_{\mu
\nu  \rho}, \quad {\cal R}G_{\mu 10} =- G_{\mu 10},
\end{equation}
$$
{\cal R}G_{10 10} = G_{10 10}, \quad  {\cal R}C_{\mu \nu 10} = C_{\mu \nu 10},
\quad \mu, \nu, \rho = 0, \dots 9,
$$
and for fermions
\begin{equation} \label{FP}
{\cal R}\psi_{\mu \alpha} = \psi_{\mu \alpha}, \quad {\cal R}\psi_{\mu {\dot
\alpha}} = - \psi_{\mu {\dot \alpha}}, \quad {\cal R}\psi_{10 \alpha} =-
\psi_{\mu 10}, \quad
  {\cal R}\psi_{10  {\dot  \alpha}} = \psi_{10 {\dot \alpha}}.
\end{equation}
At the orbifold fixed planes only fields which are even under the
action of the parity group survive, and therefore we have a field
content of $N=1$ $D=10$ supergravity i.e., bosonic fields $G_{\mu
\nu}, C_{\mu \nu 10} \equiv B_{\mu \nu}, G_{10 10} \equiv \phi$ and
fermionic fields $\psi_{\mu \alpha},\psi_{{\dot \alpha}} $. However,
since the theory is chiral we need  extra fermionic degrees of
freedom to cancel the anomaly. This is achieved by adding  two Yang
-- Mills multiplets $A_\mu^{1,2}, \xi^{1,2}_\alpha$ where each
multiplet is located on different orbifold planes and belongs to
adjoint representations of different $E_8$ gauge groups.

\subsection{$E_8$ Level Two Heterotic String Model}\label{e82}

 Let us consider in some detail the spectrum of $E_8$  heterotic string theory
\cite{Kawai:1986vd}--\cite{Dixon:1986iz} (see also \cite{Font:1990uw}--
\cite{Forgacs:1988iw})
realized at the  Kac--Moody level two.
This model can be obtained from the $E_8 \otimes E_8$ heterotic superstring by
orbifolding it
with the operator $ {\cal M}={\cal P} \otimes {\cal Q}$. Here the operator
${\cal P}=
\exp{(2 \pi i J_{12})}$ is the rotation operator.
With respect to this operator bosons have eigenvalue $+1$ and fermions have
eigenvalue  $-1$.
The operator ${\cal Q}$  interchanges the gauge degrees of freedom,
which we denote as $F^{I}$ and $\tilde F^I$.

The spectrum
of the usual $E_8 \otimes E_8$ heterotic string
 is generated by oscillators which belong to the left and right
moving sectors. In the right moving sector we have the oscillators
of the ordinary closed superstring,
\begin{equation}
[\alpha_m^\mu, \alpha_n^\nu] =-m \eta^{\mu \nu} \delta_{mn}, \quad
\{ d_m^\mu, d_n^\nu \}=-\eta^{\mu \nu} \delta_{mn}, \quad \{
b_m^\mu, b_n^\nu \}=-\eta^{\mu \nu} \delta_{mn}~.
\end{equation}
The mass formulas have the following form in the Ramond sector,
\begin{equation}
\frac{1}{4}M_R =  \sum_{n=1}^{ \infty}  \alpha^i_{-n}  \alpha^i_n +
\sum_{n=1}^{ \infty} n d^i_{-n}d^i_n~,
\end{equation}
and in the Neveu--Schwarz sector,
\begin{equation}
\frac{1}{4}M_R =  \sum_{n=1}^{ \infty}  \alpha^i_{-n}  \alpha^i_n +
\sum_{r=1/2}^{ \infty} r b^i_{-r}b^i_r - \frac{1}{2}~.
 \end{equation}
In the left moving sector we have oscillators,
\begin{equation}
[ \tilde \alpha_m^\mu, \tilde \alpha_n^\nu] =-m \eta^{\mu \nu}
\delta_{mn}, \quad [ \tilde \alpha_m^I, \tilde \alpha_n^J] =-m
\delta^{IJ} \delta_{mn}~,
\end{equation}
where $I,J = 1,..,16$ and the corresponding
 mass formula,
\begin{equation}
\frac{1}{4}  M_L =  \sum_{n=1}^{ \infty}  (\tilde \alpha^i_{-n}
\tilde\alpha^I_n +\tilde\alpha^I_{-n}  \tilde\alpha^i_n)+
\sum_{I=1}^{ 16} {(p_L^I)}^2 -1.
 \end{equation}

After orbifolding by  the operator ${\cal M}$ one obtains the
following massless spectrum: in the untwisted sector we have
 fields $G_{\mu \nu}, B_{\mu \nu}, \phi$ generated via
 \begin{equation}
b^i_{-1/2}|0  \rangle_R \otimes  \tilde  \alpha^j_{-1/2} |0  \rangle_L,
\end{equation}
since $J_{12}b^i_{-1/2}|0  \rangle_R=b^i_{-1/2}|0  \rangle_R $,
$J_{12}\alpha^j_{-1/2} |0  \rangle_L=\alpha^j_{-1/2} |0  \rangle_L$
and there are no gauge factors. The corresponding superpartners when
the right vacuum is in the Ramond sector
 \begin{equation}
|0  \rangle_R \otimes  \tilde  \alpha^j_{-1/2} |0  \rangle_L
\end{equation}
are projected out since they are not invariant under
${\cal P} \otimes {\cal Q}$
(since $J_{12}|0  \rangle_R=-|0  \rangle_R $ in the Ramond sector).

There are some more states in the untwisted sector, namely
space time bosons from the right sector should combine with the states in the
left sector which are symmetric with respect to
${\cal Q}$, whereas the space-time fermions combine with the states which are
antisymmetric with respect to ${\cal Q}$.
The corresponding states are  denoted by
\begin{equation}
b^i_{-1/2}|0  \rangle_R \otimes  |p^I  \rangle_L
\end{equation}
and
\begin{equation}
|0  \rangle_R \otimes  |\tilde p^I  \rangle_L.
\end{equation}
Thus we have $248$ gauge bosons and fermions
$A_\mu^i, \xi^i_{\alpha}$.
They belong to the adjoint representation
of the $E_8$ group realized at the Kac--Moody  level $2$.
This fact
can be seen from the following
considerations \cite{Font:1990uw},
\cite{Dienes:1996yh}--\cite{Aldazabal:1995cf}.
In the heterotic string theory Cartan generators of  $E_8 \otimes E_8$ group
are realized in terms of
$$H_1^I = \frac{i}{\sqrt 2} \partial F^i ~~{\rm and}~~
H_1^I = \frac{i}{\sqrt 2}\partial F^i.$$
>From the OPE
$$F(z)F(w) = - ln {|z-w|}^2$$
one can obtain
$$H_i^I(z) H_j^J(w) = \frac{1/2 \delta_{ij} \delta^{IJ}}{{(z-w)}^2}.$$
The later relation means that the gauge group is realized at  level
$1$. However if we take as a diagonal subgroup $H^I = H^I_1+H^I_2 $
and $E^a= E^a_1+E^a_2$ then computing the OPE between Cartan
generators we obtain $$H_i(z) H_j(w) = \frac{ \delta_{IJ}
}{{(z-w)}^2}~,$$ which means that the gauge group is realized at
level two Kac--Moody algebra. This in turn means that the roots of
the algebra became shorter in accordance with the mass formula and
the states just obtained are massless.

As is well known, in order to obtain a modular invariant
partition function, one has to add the twisted sector to the model,
{\it i.e.} consider the conditions
 \begin{eqnarray}
F^I( \sigma +  \pi - t)& = & \tilde F^I( \sigma - t) +  \pi W^I ~, \\
\nonumber
 \tilde F^I( \sigma +  \pi - t)& = & \tilde F^I( \sigma    - t) +  \pi  \tilde
W^I ~, \end{eqnarray}
where $W^I$ and $\tilde W^I$ are  vectors in
the $E_8$ lattice. The solution of these equations
 \begin{eqnarray}
F^I( \sigma - t)& = & F^I_0 + M^I(\sigma - t) + \sum_r^\infty
\frac{f_r^I}{r}e^{-2ir(\sigma-t)}   \\  \nonumber
\tilde F^I( \sigma - t)& = & \tilde F^I_0 + \tilde M^I(\sigma - t) +
\sum_r^\infty \frac{\tilde f^I_r}{r}e^{-2ir(\sigma-t)}
 \end{eqnarray}
implies the relation between oscillator modes,
\begin{equation}
f^I_r = e^{2 \pi i r} \tilde f_r^I =e^{4 \pi i r}  f_r^I~,
\end{equation}
which in turn means that $r$ can be either integer or half--integer.
The quantization leads in the massless twisted sector to the fermions
$\tilde \xi_{\dot
\alpha}$ with the opposite chirality
(being in the adjoint representation of $E_8$)
to the ones in the untwisted sector  and to a tachyon.\footnote{
This can be easily seen from mass formulae in the twisted sector for
the left movers $\frac{M^2_L}{8}= N^2 + p^2 - \frac{1}{2}$ where $N$
can take integer and half-integer values and for the right movers in
the Ramond sector $\frac{M^2_R}{8} = N$  and in the Neveu--Schwarz
sector $\frac{M^2_R}{8}= N - \frac{1}{2}$ with $N$ integer.}

\subsection{$SO(16)\otimes SO(16)$ Heterotic String Model}

Let us discuss now the case of $SO(16)\otimes SO(16)$ heterotic
string theory \cite{Kawai:1986vd}--\cite{Dixon:1986iz}. The model
can be obtained from the $E_8 \otimes E_8$ supersymmetric heterotic
string theory by orbifolding it with the operator ${\cal F}= {\cal
P} \otimes {\cal Q}$. Here again the operator ${\cal P}= \exp{(2 \pi
i J_{12})}$ is the rotation operator, while the operator ${\cal Q}$
is an element of the group $E_8\otimes E_8$ with the property
${({\cal Q})}^2=1$. This element has the form ${\cal Q}={\cal Q}_1
\otimes {\cal Q}_2$ where the element ${\cal Q}_1$ belongs to the
centre of the group $SO(16)$, which is the subgroup of $E_8$. Under
the action of this element the representation $120$ of the group
$SO(16)$ is even, while the representation  $128$ is odd. The
element ${\cal Q}_2$ has completely the  same properties with
respect to the second group $SO(16)$. The spectrum of the theory
contains the following massless fields: in the untwisted sector
there are bosonic fields  from the $N=1$ ten dimensional
supergravity multiplet $G_{\mu \nu}, B_{\mu \nu}, \phi$ and a gauge
boson $A_\mu$ which belongs to the adjoint representation $(120,1)
\oplus (1,120)$ of the  group $SO(16) \otimes SO(16)$. In the
fermionic sector one has fermions $\psi_{\alpha}$
 which belong to the
 $(128,1) \oplus (1,128)$ representation
of the  group $SO(16) \otimes SO(16)$.
In the massless part of the  twisted sector
one has just  fermions $\tilde \xi_{\dot\alpha}$ which belong to the
$(16,16) $ representation.
The model is nonsupersymmetric and chiral but anomaly free,
since the number of the fermions with the opposite chirality is the same.

\setcounter{equation}0\section{Models}

\subsection{Model 1}

In this section we explore the possibility of obtaining from the
Horava--Witten model
the massless
spectrum of the ten dimensional $E_8$ level two
heterotic string model, described in section \ref{e82}.
A natural way to achieve this
goal is to require the invariance of the spectrum of the
Horava--Witten model with respect to some extra operation, which
acts on the gauge degrees of freedom. As  discussed in  subsection
\ref{e82} the operation of interchanging of the gauge degrees of freedom
of a Lie group $G \otimes G$ breaks the gauge symmetry of the model
to the diagonal subgroup $G$ realized at level two Kac--Moody algebra.
Besides that the operation should identify the orbifold fixed planes in
order to obtain the desired gauge symmetry on a single brane.

Following these considerations
 let us introduce  two  operators ${\cal M}$ and ${\cal N}$
and require the invariance of the spectrum  of the
Horava--Witten
model under their action.
The operator
${\cal M}=  {\cal P} \otimes {\cal Q} $
is the one discussed in  subsection \ref{e82}
while the operator
${\cal N}$
acts only on the fields
at the orbifold fixed planes and
has the following properties:
\begin{eqnarray} \label{NG1} \nonumber
{\cal N}G_{\mu \nu}(x^\mu, x^{10}=0) &=&
G_{\mu \nu}(x^\mu, x^{10} =\pi), \\  \nonumber
{\cal N}B_{\mu \nu }(x^\mu, x^{10}=0) &=&
B_{\mu \nu  }(x^\mu, x^{10}=\pi), \\  \nonumber
 {\cal N } \phi(x^\mu, x^{10}=0)  &=&  \phi(x^\mu, x^{10} = \pi), \\
{\cal N}G_{\mu \nu}(x^\mu, x^{10}=\pi) &=&
G_{\mu \nu}(x^\mu, x^{10} =0), \\  \nonumber
{\cal N}B_{\mu \nu }(x^\mu, x^{10}=\pi) &=&
B_{\mu \nu  }(x^\mu, x^{10}=0), \\  \nonumber
{\cal N } \phi(x^\mu, x^{10}=\pi)  &=&  \phi(x^\mu, x^{10} =0),
\end{eqnarray}
\begin{eqnarray}
{\cal N}\psi_{\mu \alpha}(x^\mu, x^{10}=0)&=& \psi_{\mu
\alpha}(x^\mu, x^{10}= \pi),  \\  \nonumber
  {\cal N}\psi_{ {\dot  \alpha}}(x^\mu, x^{10}=0)& =& \psi_{ {\dot
\alpha}}(x^\mu, x^{10}= \pi) \\ \nonumber
{\cal N}\psi_{\mu
\alpha}(x^\mu, x^{10}= \pi)&=& \psi_{\mu \alpha}(x^\mu, x^{10}= 0),
\\  \nonumber
  {\cal N}\psi_{ {\dot  \alpha}}(x^\mu, x^{10}=\pi) &=& \psi_{ {\dot
\alpha}}(x^\mu, x^{10}= 0).
\end{eqnarray}
\begin{eqnarray} \label{NG3} \nonumber
{\cal N}A^1_{\mu }(x^\mu, x^{10}=0) &=&
A^2_{\mu } (x^\mu, x^{10}= \pi) ,   \\  \nonumber
{\cal N } \xi_\alpha^1(x^\mu, x^{10}=0) &=&
\xi^2_\alpha(x^\mu, x^{10}= \pi) ,\\
{\cal N}A^2_{\mu }(x^\mu, x^{10}=\pi) &=&
A^1_{\mu } (x^\mu, x^{10}= 0) ,   \\  \nonumber
{\cal N } \xi_\alpha^2(x^\mu, x^{10}=\pi) &=& \xi^1_\alpha(x^\mu, x^{10}= 0) ,
\end{eqnarray}
  The requirement of the invariance  of the spectrum of the
Horava--Witten model under the action of operators ${\cal M}$ and
${\cal N}$
 leads
to the following massless fields. First consider
  the case of gauge bosons
$A^{1,I}_{\mu }(x^\mu, x^{10}=0)J^{I}$ and $A^{2,I}_{\mu }(x^\mu, x^{10}=\pi)
\tilde J^I$, where
$J^I$ and $\tilde J^I$ are generators of the first and the second $E_8$ groups
respectively.
Making use  of the operator ${\cal N}$  we can introduce a combination
\begin{equation}
{\cal N}A^{1,I}_{\mu }(x^\mu, x^{10}=0)J^{I} + A^{2,I}_{\mu }(x^\mu,
x^{10}=\pi) \tilde J^I =A^{2,I}_{\mu }(x^\mu, x^{10}=\pi)(J^I + \tilde J^I).
\end{equation}
The right hand side is obviously invariant with respect to the
operator ${\cal M}$. A requirement of the invariance with respect to
the operator ${\cal N}$ gives the further constraint $A^{1,I}_{\mu
}(x^\mu, x^{10}=0)=A^{2,I}_{\mu }(x^\mu, x^{10}=\pi)$. Therefore,
one obtains a single gauge field on a brane (two initial orbifold
planes are effectively identified) and the $E_8$ gauge theory is
realized at level two  Kac--Moody algebra. For the fields $G_{\mu
\nu}, B_{\mu \nu}$ and $\phi$ the discussion is completely the same,
namely after imposing the invariance under the operators just
introduced only the symmetric combination of these fields survives.

For the fermionic fields from the Yang--Mills supermultiplet the
situation is again similar with  a slight modification. Namely,
since they are odd with respect to the operator ${\cal J}$,  the
combination which is invariant under the action of ${\cal M}$
 has the form
\begin{equation}
{\cal N}\xi^{1,I}_{\alpha}(x^\mu, x^{10}=0)J^{I} - \xi^{2,I}_{\alpha}(x^\mu,
x^{10}=\pi) \tilde J^I =\xi^{2,I}_{\alpha}(x^\mu, x^{10}=\pi)
(J^I - \tilde J^I).
\end{equation}
Finally the fermions $\psi_{\mu \alpha},\psi_{{\dot \alpha}} $ from
the ten dimensional supergravity multiplet are projected out, since
they are not invariant under the action of the operator ${\cal M}$.
Therefore requiring the invariance of the spectrum under the action
of the operators $ {\cal M}$ and  $ {\cal N}$ projects out
the fermionic sector of $10D$ Supergravity while the
rest of the fields give
rise to the untwisted sector
of the $E_8$ model.

The theory as it stands is anomalous since it is  chiral.
In order to cancel the anomaly one has to add the same number of fermions but
with the opposite chirality, let us denote them by
$\tilde \xi_{{\dot \alpha}}$.
 Since the orbifold fixed planes are now effectively identified
we need to impose the ``proper`` transformation properties with
respect to the operator ${\cal M}$. This means that the fermions
belong to the adjoint representation of the gauge group $E_8$ and
have the form $ \tilde \xi^I_{\dot \alpha}(J^I - \tilde J^I)$. These
fermions are analogous to those which appear in the
twisted sector of the
$E_8$ level two heterotic string model, thus leading to an absence of
anomalies and to its modular invariance.

Finally, let us note that the ten dimensional $E_8$ heterotic
string theory is tachyonic whereas the eleven dimensional orbifold
that we considered is nonsupersymmetric and tachyon free. One
possibility is to simply add a tachyonic field to the spectrum. A
more interesting possibility is a case when the tachyon becomes
massive at the limit we are considering. We note that such
decoupling of the tachyon as a function of moduli is a well known
phenomenon in string vacua
\cite{Ginsparg:1986wr} --
\cite{Angelantonj:2006ut}.

\subsection{Model 2}

Another possibility is to consider a model when one
keeps the entire supergravity multiplet intact.
For this reason one has to assign  nontrivial transformation properties
under  the operator ${\cal N}$
 to
the fermions $\psi_{\mu \alpha},\psi_{{\dot \alpha}} $ from the
$N=1$ ten dimensional supergravity multiplet in analogy with
(\ref{NG1})--(\ref{NG3}). However, in order to keep these fermions
unprojected this is not enough and one has to modify the form of the
operator ${\cal M}$ as well, since all fermionic degrees of freedom
in the initial Horava--Witten model were odd with respect to the
operator
 ${\cal J}$.
The simplest choice is to  take ${\cal M}= {\cal Q}$ for this model.
The transformation properties under the action of ${\cal N}$ for the
other fields are again (\ref{NG1}) -- (\ref{NG3}). The discussion of
the spectrum is now completely analogous to the previous model.
Namely in the Yang--Mills sector of the model we again have a ten
dimensional Yang--Mills supermultiplet
 with the gauge group  $E_8$
realized at the Kac--Moody level two and its fermionic part is
obtained by taking
\begin{equation}
{\cal N}\xi^{1,I}_{\alpha}(x^\mu, x^{10}=0)J^{I} + \xi^{2,I}_{\alpha}(x^\mu,
x^{10}=\pi) \tilde J^I =\xi^{2,I}_{\alpha}
(x^\mu, x^{10}=\pi)(J^I + \tilde J^I).
\end{equation}
On the supergravity side we have a complete  ten dimensional $N=1$
supergravity multiplet. As in the previous model, the model is
chiral since, though the supergravity multiplet has been kept, the
number of fermionic degrees of freedom in the super Yang--Mills
multiplet has been halved. Therefore, in order to cancel the
anomalies, one has to add extra fermions $ \tilde \xi^I_{
\alpha}(J^I + \tilde J^I)$ with the same chirality as the ones from
the Yang--Mills supermultiplet, in the adjoint representation of the
gauge group.

To summarise, this model is quite similar to the Horava--Witten
model but with
the gauge symmetry realized at the Kac--Moody level
two.

\subsection{Model 3}

Finally let us discuss how one can obtain the massless spectrum of
the nontachyonic $SO(16)\otimes SO(16)$ heterotic string  from the
Horava--Witten model by imposing on it some extra symmetries.
Similarly to the previous examples this extra symmetry actually
coincides with the one which is imposed  on the $E_8 \otimes E_8$
heterotic string to obtain the ten dimensional $SO(16)\otimes
SO(16)$ model, but enhanced by the extra operator ${\cal N}$.
Therefore, the symmetry under consideration is generated by operators
${\cal F}$ and
${\cal N}$ where the operator ${\cal F} $ is defined in subsection
2.3. Requiring invariance of the spectrum of the
Horava--Witten model with respect to the action of
${\cal F}$ projects out the fermions
 $\psi_{\mu \alpha},\psi_{{\dot \alpha}} $ from the
 supergravity spectrum and leads to the massless fields
 which belong to the massless untwisted sector of the
 ten dimensional  $SO(16)\otimes SO(16)$ model.
The action of the operator ${\cal N}$ on the bosonic fields  from
the ten dimensional supergravity multiplet is the same as
(\ref{NG1}), while for the field from the Yang--Mills supermultiplet
it reads
\begin{eqnarray} \label{NG5} \nonumber
{\cal N}A^1_{\mu }(x^\mu, x^{10}=0) &=& A^1_{\mu }
(x^\mu, x^{10}= \pi) ,   \\  \nonumber
{\cal N } \xi_\alpha^1(x^\mu, x^{10}=0)  &=&
\xi^1_\alpha(x^\mu, x^{10}= \pi) ,\\
{\cal N}A^2_{\mu }(x^\mu, x^{10}=\pi) &=& A
^2_{\mu } (x^\mu, x^{10}= 0) ,   \\  \nonumber
{\cal N } \xi_\alpha^2(x^\mu, x^{10}=\pi)  &=&
\xi^2_\alpha(x^\mu, x^{10}= 0) ,
\end{eqnarray}
which simply means that we are putting the orbifold branes on top of
each other. This model is again anomalous, since as in the previous
models we are missing massless fermions from the twisted sector.
Therefore, to cancel the anomalies we add an extra set of fermions
$\tilde \xi_{ \dot  \alpha}$ with the opposite chirality to the ones
of the fermions from the Yang--Mills supermultiplet. We can take
these fermions to be in the $(16,16)$ representation of the $SO(16)
\otimes SO(16)$, gauge group and this choice leads to the massless
spectrum of the corresponding heterotic string model. This choice is
not unique, however. Namely,
 one can take  these extra  fermions to be exactly in the same representation
 $(128,1)\oplus (1,128)$
of the gauge group as those from the Yang--Mills supermultiplet.
 In this way the model will be anomaly free since it is nonchiral, but
it has no string theory counterpart.

\section{Conclusions}

String theory provides an internally consistent framework for the
phenomenological approach to quantum gravity. Important progress has
been achieved in string theory with the realization that the five
supersymmetric string theories in ten dimensions are related by
perturbative and nonperturbative duality transformations. This
understanding is, however, not complete as the non--supersymmetric
ten dimensional string theories are not incorporated in the duality
picture. It is plausible that the understanding of some
phenomenological issues in string theory is contingent on
understanding how the nonsupersymmetric string vacua fit into the
duality picture. In this note we explored the possibility of
generating the spectrum of the $E_8$ and $SO(16)\times SO(16)$ ten
dimensional heterotic string theories as orbifolds of the
Horava--Witten theory. We argued that this is indeed possible by
imposing the appropriate identification conditions on the fields
that reside on the two fixed branes. We further demonstrated the
conditions required for the resulting theories to be anomaly free.
One can  contemplate that the generated orbifolds are the
nonperturbative limits of the ten dimensional nonsupersymmetric
string theories and subject this hypothesis to further tests.

Our basic starting point is the eleven dimensional supergravity
field theory compactified on $S_1/Z_2$, \`a la Horava--Witten. The
basic point in our analysis is the gluing of the end points of the
Horava--Witten theory at $x^{10}=0$ and $x^{10}= \pi$, and the
identification of the fields present on the boundary branes, located
at the orbifold fixed points $x^{10}=0$ and $x^{10}= \frac{\pi}{2}$.
This operation is augmented by an additional operation which acts on
the gauge degrees of freedom, in analogy with the one discussed in
ref. \cite{Dixon:1986iz}. Identification of the degrees of freedom
on two different branes located at the different orbifold fixed
points entails the identification of the branes themselves in order
to maintain the locality of the low energy theory. The massless
spectrum in these models is obtained from that of the Horava--Witten
model by taking the eleventh coordinate to be compact. In other
words, the operation entails putting the orbifold fixed planes on
top of each other, with additional identification of gauge degrees
of freedom. This operation projects out certain fields from the
initial spectrum. Anomaly cancellation dictates the addition of
fermionic fields with certain chirality on the resulting brane, in
analogy with the Horava--Witten construction. We reproduced the
massless spectra of various nonsupersymmetric ten dimensional
heterotic string vacua from the Horava--Witten model, by the
operation which essentially changes the geometry of the later. We
remark that one can consider an alternative orbifold by moding the
eleventh dimension by the reflection symmetry under the exchange
$x^{10} \leftrightarrow {\pi -x^{10}}$. Under this orbifold there
would be a fixed orbifold point at $x^{10}=\pi/2$. This operation
differs from the one that we considered in this paper, and may be of
interest for further exploration.

\vspace{1cm}

\noindent {\bf Acknowledgements} \newline We would like to thank
Carlo Angelantonj, Nikos Irges,
 Elias Kiritsis and Thomas Mohaupt for discussions. AEF would like to thank the
INFN  Galileo Galilei Institute in Florence for hospitality. This
work was supported in part by UK Science and Technology Facilities
Council under the grant PP/D000416/1 and by the ``UniverseNet''
MRTN--CT--2006--035863 and ``Superstring Theory''
MRTN--CT--2004--512194.


\renewcommand{\thesection}{A}

\setcounter{equation}{0}

\renewcommand{\theequation}{A.\arabic{equation}}

\end{document}